\documentclass[aip,twocolumn,preprint,showpacs,preprintnumbers,amsmath,amssymb,superscriptaddress]{revtex4-1}

\usepackage{graphicx}
\usepackage{dcolumn}
\usepackage{bm}
\usepackage{hyperref}

\begin{document}

\title{Broadband microwave and time-domain terahertz spectroscopy of CVD grown graphene}

\author{W. Liu}
\affiliation{Department of Physics and Astronomy, The Johns
Hopkins University, Baltimore, MD 21218,USA}

\author{R. Vald\'es Aguilar}
\affiliation{Department of Physics and Astronomy, The Johns
Hopkins University, Baltimore, MD 21218,USA}

\author{Y. Hao}
\affiliation{Department of Mechanical Engineering and the Materials Science and Engineering Program, University of Texas - Austin, Austin,
TX 78712-0292, USA}

\author{R.S. Ruoff}
\affiliation{Department of Mechanical Engineering and the Materials Science and Engineering Program, University of Texas - Austin, Austin,
TX 78712-0292, USA}

\author{N.P. Armitage }
\affiliation{Department of Physics and Astronomy, The Johns
Hopkins University, Baltimore, MD 21218,USA}

\date{\today}

\begin{abstract}

We report a study of the complex AC impedance of CVD grown graphene.  We measure the explicit frequency dependence of the complex impedance and conductance over the microwave and terahertz range of frequencies using our recently developed broadband microwave Corbino and time domain terahertz spectrometers (TDTS).   We demonstrate how one may resolve a number of technical difficulties in measuring the intrinsic impedance of the graphene layer  that this frequency range presents, such as distinguishing contributions to the impedance from the substrate.  From our microwave measurements,  the AC impedance has little dependance on temperature and frequency down to liquid helium temperatures.  The small contribution to the imaginary impedance comes from either a remaining residual contribution from the substrate or a small deviation of the conductance from the Drude form.

\end{abstract}



\maketitle
\section{Introduction}
Graphene is a material consisting of a single atomic carbon layer arranged in a 2D honeycomb lattice, discovered in 1969 \cite{may69}and studied extensively since then \cite{dreyer10} by the surface science community. It has attracted wide-spread interest for both its novel electronic properties and Dirac band dispersion as well as its broad application potential \cite{berger04a,Kim05a,novoselov05,geim07a,neto09a,dreyer10,zhu10,Sarma11a}. Due to its high mobility, it also has been proposed to show great promise for high speed switching in microwave and terahertz devices \cite{rangel08,wang09,Dragoman09} and terahertz plasmon amplification \cite{rana08}.

Recently it has been demonstrated that large-area monolayer graphene films can be grown by chemical vapor deposition (CVD) on copper foils\cite{reina08a}, following the precipitation-based growth of somewhat non-uniform few-layer graphene films on Ni foils\cite{yu08a,kim09anature}.
This method \cite{li09a,li09b,li10a,li11a} allows the growth of large scale graphene films that can be transferred to various substrates, which will be essential for any practical device applications. The availability of large area uniform graphene also allows access to these materials from a greater variety of experimental techniques including studies of their long wavelength electromagnetic response. Their complex microwave and terahertz frequency dependent response are of particular interest and their understanding is crucial in order to use graphene for fast electronic devices. However, it has traditionally been difficult to get significant broadband spectral information in these frequency ranges, particularly in the microwave regime. Microwave experimental techniques are typically very narrow band and may at best allow the characterization of materials at only a few discrete frequencies.

 In this study, we make use of our newly-developed microwave ``Corbino" spectrometer to measure the broadband microwave response in the frequency range from 100 MHz to 16 GHz of CVD grown graphene at temperatures down to 330 mK.  This technique allows one to gain broadband spectral information in the microwave regime.   Microwave techniques are typically very narrow band. We present data for both the sheet impedance and complex conductance.  The measurement of the intrinsic impedance of a single atomic layer film on an insulating substrate presents a number of experimental challenges in the microwave regime.  As a non-resonant technique, the Corbino spectrometer requires an intricate calibration procedure.  More importantly, any attempt to measure the intrinsic impedance of graphene will be affected by capacitive coupling to its dielectric substrate.   At microwave frequencies, the impedance associated with the substrate can be a substantial fraction of the graphene impedance.   We detail the manner in which substrate effects may be calibrated for, as well as a number of other difficulties peculiar to this frequency range that must be overcome.   We also compare our data to that we measure at higher frequencies using time-domain terahertz spectroscopy.  Both measurements techniques are capable of measuring the complex optical response functions as a function of temperature and frequency, without resorting to Kramers-Kronig transforms.

\begin{figure}[h]
\begin{center}
\includegraphics[width= 0.85\columnwidth,angle=0]{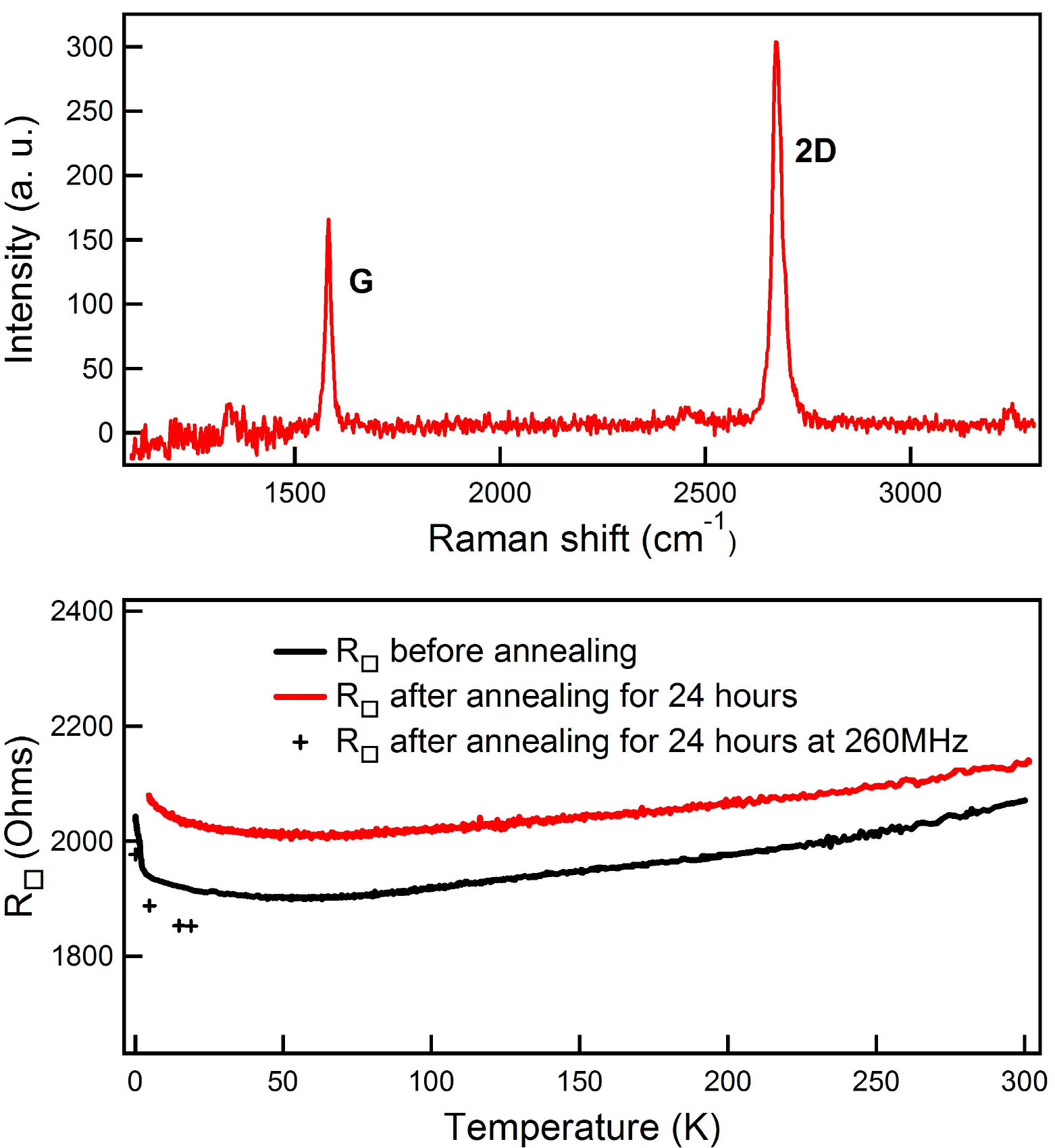}
\caption{ (Color Online) (a)  Raman spectra of graphene on Si. The Raman signal from a bare Si substrate has been subtracted. (b)  Resistance per square R$_\Box$ for one of the CVD grown graphene samples as a function of temperature. The black crosses are the microwave data at 260 MHz. After the same sample was annealed at about 60 C in vacuum for about 24 hours, the temperature dependence of its resistance shifted as shown by the red curve.}
\label{ramanandDC}
\end{center}
\end{figure}

 CVD grown Graphene was prepared by methane and hydrogen at pressures of 1.5 mbar over a 25 $\mu$m thick Cu foil.  The graphene films are coated
with PMMA and then the Cu foils are dissolved in an aqueous solution of FeCl$_{3}$.  The graphene is rinsed several times with de-ionized water and can then be scooped out of solution onto a 380 $\mu$m thick clean high resistivity Si substrate.  High purity Si was used as it has a purely dielectric contribution to the impedance of the graphene-silicon multilayer.   The sample is allowed to dry and adhere to the Si and then the PMMA is removed by acetone.  The resulting films are verified to be of high-quality, predominantly single layer graphene from the intensities and positions of the G- and 2D-band peaks in their 532 nm Raman spectra \cite{ferrari06} (Fig. \ref{ramanandDC} (a)).

\section{Microwave Measurements}
The microwave `Corbino' technique has been developed to provide the frequency dependent response in this spectral range, where broadband measurements have been traditionally been very challenging.   It has been used in recent years to give important information about high temperature superconductors, heavy-fermion, electron glasses, and thin film superconductors \cite{booth96,SchefflerNature05,Lee90a,KitanoPRB09a,liu11a}.  In this technique one measures the complex reflection coefficients $S_{11}^m$  from a thin film sample that terminates a coaxial transmission line.  Typically, electrical contact is made by pressing a modified microwave adapter against the sample with donut shaped gold pads, which matches the radial dimensions of the coaxial line.  The technique necessitates the use of three calibration samples with known reflection coefficients (20nm NiCr on Si, a blank Si substrate and a bulk copper sample) to remove the effects of extraneous reflections, phase shifts and losses in the coaxial cables  \cite{Scheffler04a}.  Using the telegraphers' equation, the actual reflection coefficient at the sample surface $S_{11}^a$ is expressed as:

\begin{equation}
S_{11}^a = \frac{S_{11}^m - E_D }{E_R + E_S (S_{11}^m - E_D)}.
\label{errorcoeff}
\end{equation}

\noindent Here, the complex error coefficients  $E_D$, $E_S$ and $E_R$ represent the so-called directivity, source match and reflection tracking errors respectively. They are temperature and frequency dependent complex coefficients.  To extract out the sample sheet impedance $Z_S^{eff}$, in principle the standard equation

\begin{equation}
Z_S^{eff}= g \frac{1+S_{11}^a}{1-S_{11}^a}Z_{0}
\label{ReflectionEq}
\end{equation}

\noindent may be used.  Here $Z_{0}=$50 ohms is the characteristic cable impedance and $g = 2\pi/\ln(r_{2}/r_{1})$  is a geometric factor where $r_{2}$ and $r_{1}$ are the outer and inner radii of the donut shaped sample. However additional complications are presented in the case of materials like graphene where the impedance of the film is comparable to that of the substrate, as the substrate contribution must be taken into account explicitly. In the thin film approximation and under the assumption that only TEM waves propagate in the transmission lines, the effective impedance for a thin film of impedance $Z_{S}$ backed by a substrate with characteristic impedance $Z_S^{Sub}$  can be shown to be\cite{booth96thesis}

\begin{equation}
Z^{eff}_S = \frac{Z_S }{1+\frac{Z_S}{Z^{Sub}_S}}.
\label{subcorr}
\end{equation}

To isolate the impedance of the graphene layer, we use Eq. \ref{subcorr} with $Z_S^{Sub}$ from our previous independent study of thin superconducting films on identical Si substrates \cite{liu11a}.  In that study, we used amorphous metal films with scattering rates ($\approx$ 100 THz) so high that the intrinsic AC impedance of the film itself in the normal state was purely real and could be deduced from the DC resistance exactly.  Thus, $Z_S^{Sub}$ can be calculated by comparing the measured AC impedance of the metal film with its known value.  Knowing  $Z_S^{Sub}$ one can isolate the intrinsic $broadband$ impedance of the graphene layer.

\begin{figure}[h]
\begin{center}
\includegraphics[width=0.85\columnwidth,angle=0]{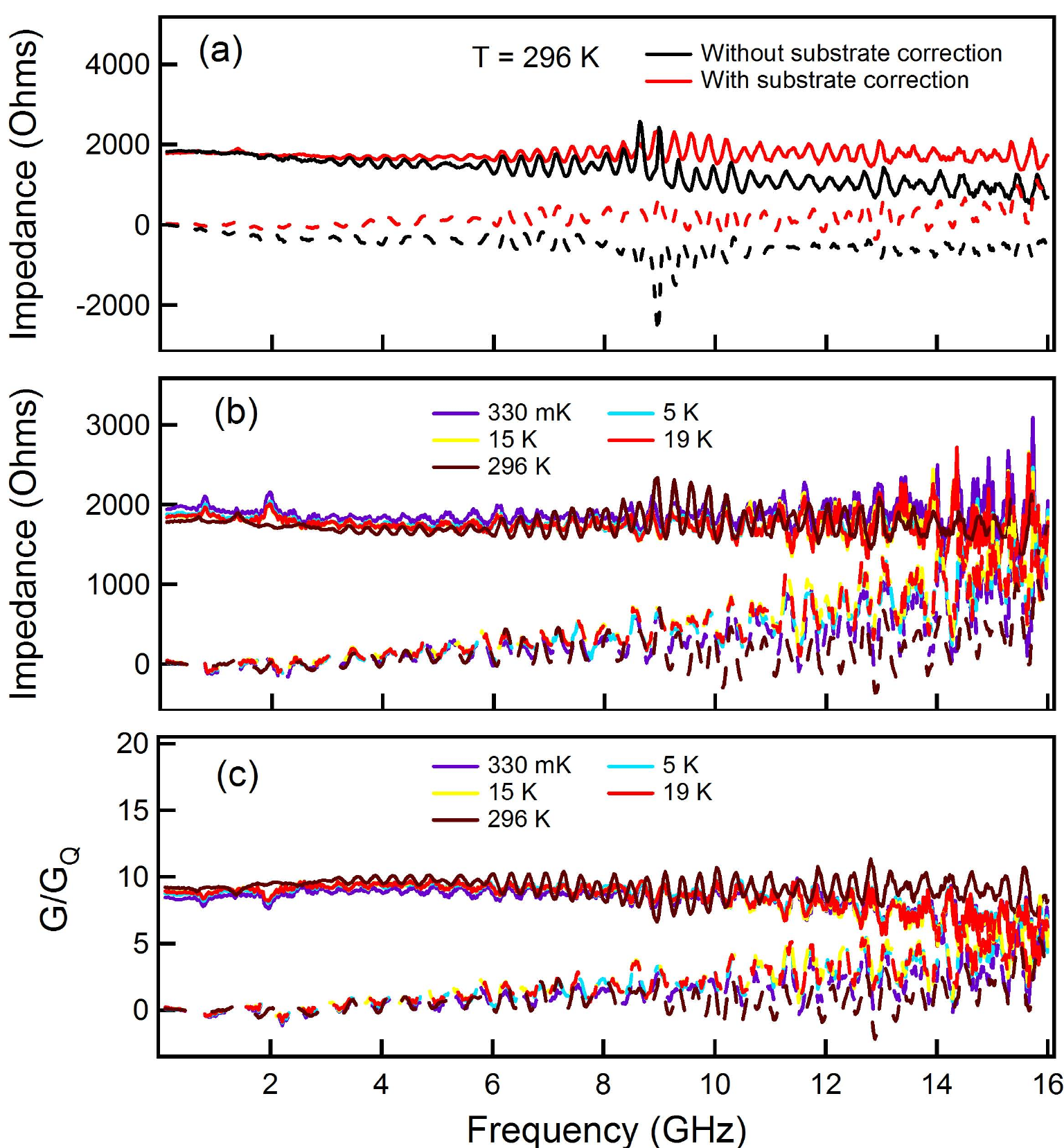}
\caption{(Color Online) (a) Calibrated impedance with (black) or without (red) correction for the substrate contribution at room temperature. Corrected (b) impedance and (c) conductance  (normalized by $G_Q= \pi e^2/2h$) as a function of frequency in the range of 100 MHz to 16 GHz at different temperatures. Different colors in both panels indicate different temperatures. In all the panels, solid lines are the real parts while dashed lines are for the imaginary parts.}
\label{impedandcond}
\end{center}
\end{figure}

In addition to measurements at microwave frequencies, the Corbino spectrometer system can measure the two contact DC resistance simultaneously using a lock-in amplifier and a bias tee.  Multiplying the measured resistance between the inner and outer conductors of the coaxial cable by the geometric factor \textit{g}, we obtain resistance per square.   As shown in Fig. \ref{ramanandDC} (b),  the resistance per square as a function of temperature is approximately temperature independent ($\approx 3\%$ over the range) with only an upturn below 30K as the principal distinguishing feature.   The sample properties are changed only slightly by annealing in vacuum for about 24 hours at 60 C degree.   As shown in Fig. \ref{ramanandDC}(b) the resistance per square increased by 5 \% with almost the same temperature dependence.  We ascribe the change in the overall scale due to a change in carrier density by driving off absorbed gases.

 In Fig. \ref{impedandcond}, we present the results of our broadband microwave measurements on one particular graphene sample from 100 MHz to 16 GHz at temperatures down to 330 mK.   The small oscillations are the residual effects of standing wave resonances in the transmission line that have been imperfectly removed by the calibration procedure.  In Fig. \ref{impedandcond}(a), we compare the effective impedance at the sample surface calculated from Eq. \ref{ReflectionEq} with the impedance of the sample after the substrate correction described above.   One can see that the correction becomes significant at higher frequency where the effective capacitance of the dielectric substrate plays a larger role.  After correction the impedance becomes primarily real and frequency independent as expected for a conductor with a scattering rate larger than the measurement range.  One can see that it is essential to perform such a correction to quantify the impedance correctly.

In Fig. \ref{impedandcond} (b), we plot both real ($Z_1$) and imaginary ($Z_2$) sample impedance corrected for the substrate contribution as a function of frequency at 5 different temperatures. We can see that the frequency dependance of impedance at the base temperature of 330 mK and room temperature are almost the same. The vertical difference in those temperatures matches with the difference of measured DC resistivity. As clearly seen from Fig. \ref{impedandcond} (b), the real and imaginary parts of impedance have little dependence on frequency down to low temperatures. This can also be seen in Fig. \ref{ramanandDC} (b), where the resistance at the low frequency of 260 MHz at 4 different temperatures is also plotted.  These data follow the DC values closely indicating a consistency of AC and DC measurements. Also, the real and imaginary parts of conductance  have little dependance on temperature which means that the Drude response of electrons in graphene does not change a lot over this wide range of temperatures (from room temperature to 330 mK). It also indicates that  the scattering rate  $\tau$  - the average time between scattering events -  bears little dependance in temperature.
\begin{figure}[h]
\begin{center}
\includegraphics[width= 0.85\columnwidth,angle=0]{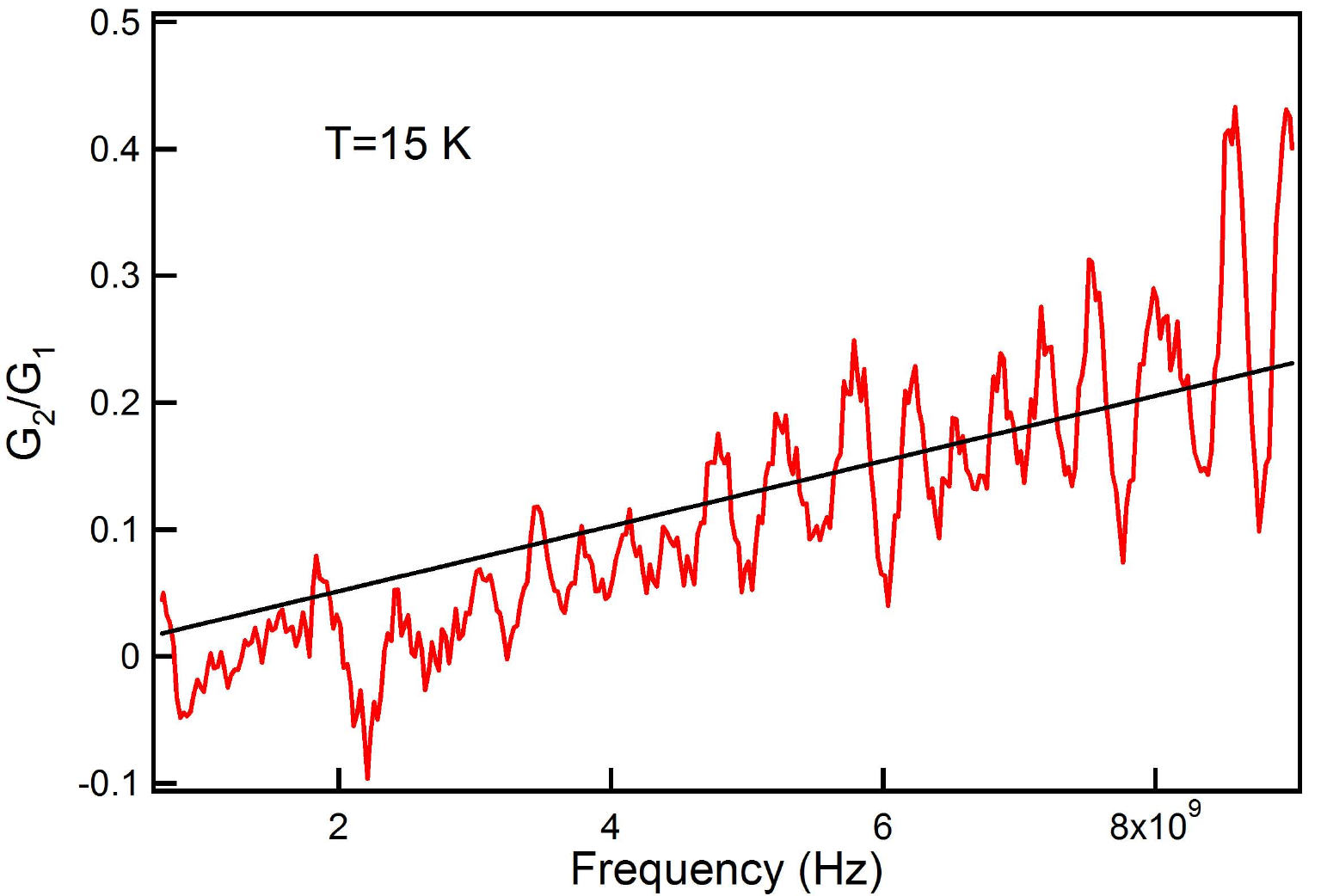}
\caption{(Color Online) Ratio (red) of real and imaginary parts of conductance  as a function of frequency in the range from 700 MHz to 9 GHz at 15 K before annealing.  The black curve is a linear fit with zero $y$ axis intercept. }
\label{tau}
\end{center}
\end{figure}

In Fig. \ref{impedandcond} (c),  we plot the complex conductance obtained by using the inverse of the data Fig. \ref{impedandcond} (b). Since in the thin film limit, the complex conductance is the reciprocal of complex impedance, it also has almost no dependance on frequency.  Here we have ratioed this data to the quantum of conductance $G_Q = \pi e^2/2h$ = 1 / 16433 Ohms$^{-1}$ expected for a graphene sample with its chemical potential tuned to the Dirac point.  Its large dimensionless scale shows that the carrier density for this sample is high with the chemical potential far from the Dirac point.

Although the imaginary part of the conductivity is very small in Fig. 2, it is not zero, which gives a measurable $\tau$  from the data.  Within the Drude model, the complex conductivity from a charge responding to a time varying external oscillating electromagnetic field with frequency $\omega$ is $\sigma(\omega) = \frac{ne^2\tau}{m} \frac{1}{1 - i \omega \tau}$  where $n$ is the electron density. Inspection of this equation shows that the ratio of the imaginary to real parts of $\sigma_2/\sigma_1$ gives $\omega \tau$ as a function of $\omega$ such that $\tau$ can be determined from its slope.   In Fig. \ref{tau}, we plot this ratio for the $conductances$ ($G = \sigma d$) as a function of frequency at 15 K. $G_2/G_1$  at other temperatures give similar results as both $G_2$ and $G_1$ have little dependence on temperature.  The slope of $G_2/G_1$ gives us an estimate of the relaxation time, which is about 25.7 ps. A rough estimation of scattering rate is then $\Gamma = 1/\tau$ = 38.9 GHz. This is quite close to the value for the Drude scattering rate of 36.4 GHz obtained independently through fitting our data using a Drude-Lorentz model  \cite{kuzmenko05a}.
Using this extracted  scattering rate, we can estimate the mean free path $l=V_F*\tau=28.3$ $\mu$m with Fermi velocity $v_F=1.1*10^6 m/s$\cite{zhang05}. However, we should note that this small value of the scattering rate is interesting as it is at odds with that inferred from previous studies using higher frequency time-domain terahertz spectroscopy \cite{Tomaino11a} or far-infrared reflectivity measurements \cite{horng11}.


\begin{figure}[h]
\begin{center}
\includegraphics[width= 0.85\columnwidth,angle=0]{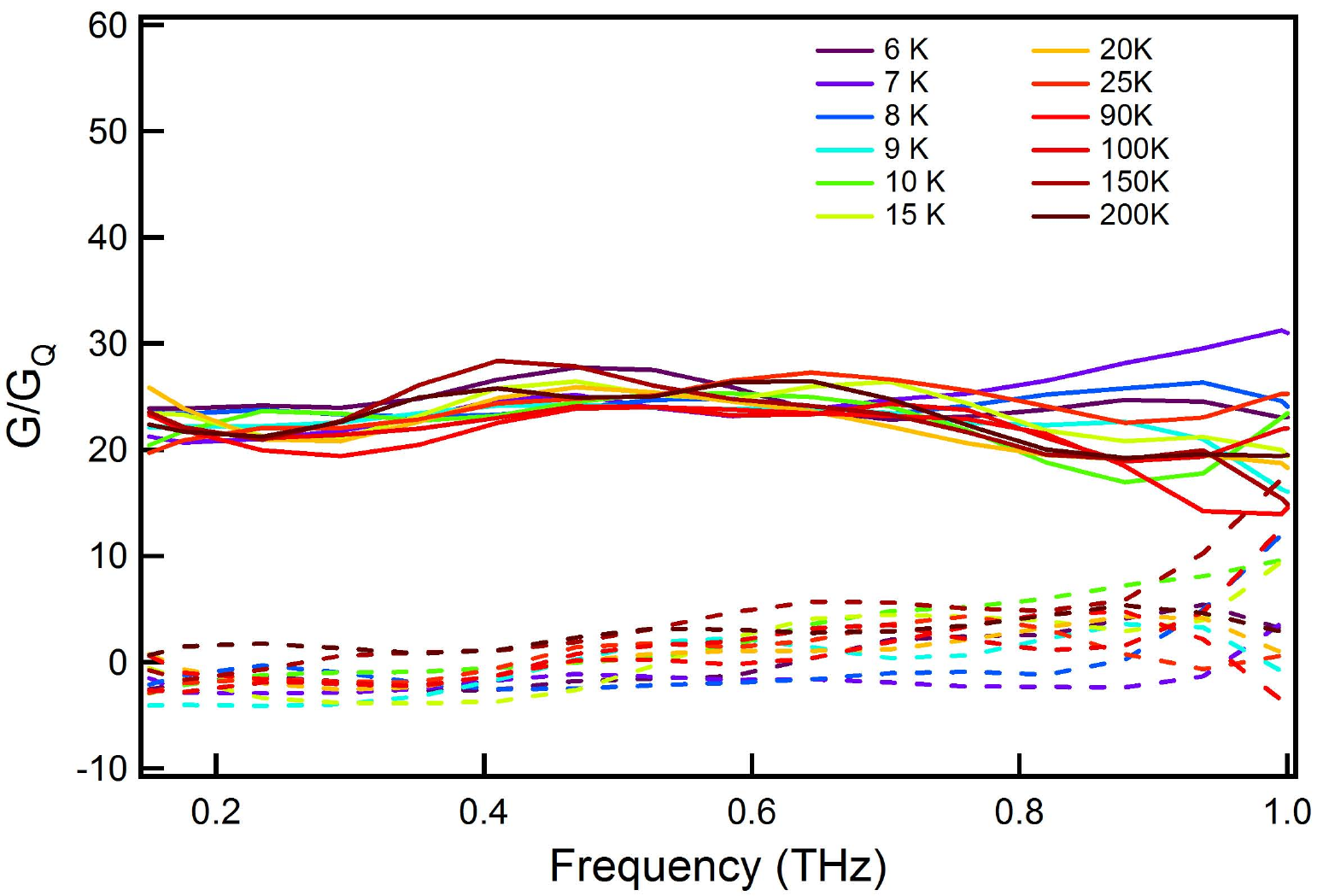}
\caption{ (Color Online) Real (solid) and imaginary (dashed) parts of normalized conductance as a function of frequency in the range from 150 GHz  to 1.0 THz at different temperatures for a similar graphene sample.}
\label{THz}
\end{center}
\end{figure}

\section{Terahertz measurements}
The small value of the scattering rate is also different from that inferred from our own time-domain terahertz  spectropscopy (TDTS) measurements on another similarly prepared sample.  In TDTS an ultrafast laser pulse excites a semiconductor switch, which generates an almost single cycle pulse with frequencies in the terahertz range.  The transmitted terahertz pulse's electric field is mapped out as a function of time. The ratio of the Fourier transform of the transmission through the sample to that of a reference (usually the substrate on which the sample is deposited) gives the complex transmission function of the film under study.  This can be inverted using standard formulas in the `thin film approximation' $\tilde{T}(\omega)=\frac{1+n}{1+n+Z_0\tilde{\sigma}(\omega)d} e^{i\Phi_s}$ to get the complex conductivity.  Here $\Phi_s$ is the phase accumulated from the small difference in thickness between the sample and reference substrates and $n$ is the substrate index of refraction.  We have measured in terahertz frequency range from 150 GHz to 1.0 THz at temperatures down to 6K with flowing He4 gas. It is possible that the He4 gas environment for graphene sample in TDTS might make a  slight difference for its scattering rate compared with the Corbino system where the sample is sealed in high vacuum.

In Fig. \ref{THz}, we show terahertz complex conductance data for a sample from a different batch.  This sample has a  higher carrier concentration as evidenced by its larger conductance. The sample also shows little dependence on frequency of the real part of the conductivity and a small imaginary conductivity in this range.   As the conductance is expected to drop off dramatically around the frequency of the scattering rate within the Drude model, this sample has a scattering rate  greater than 1.0 THz.   This is consistent with previous work \cite{Tomaino11a,horng11} on CVD grown graphene and on few-layer epitaxial graphene\cite{choi09}.

\section{conclusion}
The two different scattering rates found in our microwave and TDTS measurement  show either the limitations of the Drude model for describing the fine details of electron transport in graphene at low frequencies, or alternatively the inherent difficulties of extracting out the precise complex impedance of graphene.  In the first case it may be that the details of scattering Dirac electrons give a conductivity lineshape that is not precisely Lorentzian.   Then the slope of $G_2/G_1$ can not be taken as a measure of $\tau$.   In the second case the scattering rate may be underestimated by imprecisely removing the effects of the substrate contribution.   It may be that this overestimates the size of the imaginary contribution to the impedance and thereby giving a larger contribution to $G_2$.   Future work will attempt to measure $changes$ in the impedance by back gating the sample.   It is likely that it will be possible to measure changes in the impedance very precisely as the bias is swept from positive to negative.  Moreover, by going to the high bias regime/large conductivity regime we should be insensitive to influence of the substrate impedance and will be able to isolate the graphene conductance more precisely.

\section{ACKNOWLEDGMENTS}

The research at JHU was supported by NSF DMR-0847652.   The research at UT was supported by NSF Grant CMMI-0700107, the Nanoelectronic Research Initiative (NRI-SWAN; No. 2006-NE-1464), and the DARPA CERA grant.

\end{document}